\documentclass[12pt]{article}
\usepackage{graphicx}
\usepackage{color}
\usepackage{cite}
\usepackage{here}
\usepackage{caption}

\makeatletter
\g@addto@macro\bfseries{\boldmath}
\makeatother

\def \beq{\begin{equation}}
\def \eeq{\end{equation}}
\def\eqref#1{(\ref{#1})}
\def\bea{\begin{eqnarray}}
\def\eea{\end{eqnarray}}
\def\jpsi{J\kern-0.1em/\kern-0.1em\psi}
\def\nl{\hfill\break}

\def\URLtilde{\lower0.2em\hbox{$\tilde{\phantom{a}}$}}
\def\mycomm#1{\hfill\break\strut\kern-3em{\color{red}\tt ====> #1
\color{black}}\hfill\break}

\newcount\timecount
\newcount\hours \newcount\minutes  \newcount\temp \newcount\pmhours
\hours = \time
\divide\hours by 60
\temp = \hours
\multiply\temp by 60
\minutes = \time
\advance\minutes by -\temp
\def\hour{\the\hours}
\def\minute{\ifnum\minutes<10 0\the\minutes
\else\the\minutes\fi}
\def\clock{
\ifnum\hours=0 12:\minute\ AM
\else\ifnum\hours<12 \hour:\minute\ AM
\else\ifnum\hours=12 12:\minute\ PM
\else\ifnum\hours>12
\pmhours=\hours
\advance\pmhours by -12
\the\pmhours:\minute\ PM
\fi
\fi
\fi
\fi
}

\def\monthname{\relax\ifcase\month 0/\or January\or February\or
March\or April\or May\or June\or July\or August\or September\or
October\or November\or December\else\number\month/\fi}

\def\bold#1{\setbox0=\hbox{$#1$}     \kern-.025em\copy0\kern-\wd0
\kern.05em\copy0\kern-\wd0
\kern-.025em\raise.0433em\box0 }

\begin{document}
\setcounter{footnote}{1}
\vskip1.5cm

\begin{center}
{\large \bf \boldmath
Quark-level analogue of nuclear fusion
\\
\vrule width 0pt height 3.0ex
with doubly-heavy baryons
\unboldmath}
\end{center}
\bigskip

\centerline{Marek Karliner$^a$\footnote{{\tt marek@proton.tau.ac.il}}
 and Jonathan L. Rosner$^b$\footnote{{\tt rosner@hep.uchicago.edu}}}
\medskip

\centerline{$^a$ {\it School of Physics and Astronomy}}
\centerline{\it Raymond and Beverly Sackler Faculty of Exact Sciences}
\centerline{\it Tel Aviv University, Tel Aviv 69978, Israel}
\medskip

\centerline{$^b$ {\it Enrico Fermi Institute and Department of Physics}}
\centerline{\it University of Chicago, 5620 S. Ellis Avenue, Chicago, IL
60637, USA}
\bigskip
\strut

\begin{center}
ABSTRACT
\end{center}
\begin{quote}
The recent discovery by LHCb of the first doubly-charmed 
baryon $\Xi_{cc}^{++} = ccu$ at $3621.40 \pm 0.78$ MeV
implies a large binding energy $\sim 130$ MeV between the two $c$ quarks.
This strong binding enables a quark-rearrangement exothermic reaction
$\,\Lambda_c \Lambda_c \to \Xi_{cc}^{++}\,n\,$ with $Q=12$ MeV, which is a
quark-level analogue of 
deuterium-tritium nuclear fusion reaction $DT\to {}^4{\rm He}\,n$. 
Due to much larger binding energy between two $b$ quarks $\sim 280$ MeV,
the analogous reaction with $b$ quarks,
$\,\Lambda_b \Lambda_b \to \Xi_{bb}\,N\,$ is expected to have 
a dramatically larger $Q$-value, $138\pm 12$ MeV.

\end{quote}
\smallskip

\bigskip


Very recently LHCb has observed the doubly-charmed baryon
\,$\Xi_{cc}^{++} = ccu\,$ with a mass of $3621.40 \pm 0.78$ MeV
\cite{Aaij:2017ueg}.  This value is consistent with several predictions,
including our value of $3627 \pm 12$ MeV \cite{Karliner:2014gca,note1}.
The essential ingredient in Ref.~\cite{Karliner:2014gca} is 
the large binding energy of the two heavy quarks in a baryon,
$B(cc)=129$ MeV and $B(bb)=281$ MeV.

To a very good approximation this binding energy is 1/2 of the 
quark-antiquark binding energy in corresponding quarkonia. 
This 1/2 rule is exact in the one-gluon-exchange limit and has
now been validated by the LHCb measurement of the $\Xi_{cc}$ mass.
Its successful extension beyond weak coupling implies that the heavy quark 
potential factorizes into a color-dependent and a space-dependent
part, with the space-dependent part being the same for quark-quark and 
and quark-antiquark.  The relative factor 1/2 is then automatic, just as in
the weak-coupling limit, resulting from the color algebra.

The large binding energy between heavy quarks has some striking implications,
such as the existence of a {\em stable} $bb\bar u\bar d$ tetraquark 
with $J^P=1^+$ \cite{Karliner:2017qjm}, 215 MeV below the
$B^-\bar B^{*0}$ threshold and 170 MeV below threshold for decay to
$B^-\bar B^0 \gamma$.

In the present work we point out another striking consequence of the very
strong binding between heavy quarks: the {\em quark-level} analogue of
nuclear fusion. To start, consider the quark-rearrangement reaction
\beq
\label{eq:LcLc}
\begin{array}{ccc}
\Lambda_c\,\,\Lambda_c & \to & \Xi_{cc}^{++}\,\,n, 
\\
cud\,\,cud             &     &  ccu\,\,ddu
\end{array}
\eeq
All the masses are known and the $Q$-value is 12 MeV, as shown in Table
\ref{tab:Q}.

\begin{table}[t]
\captionsetup{singlelinecheck=off,justification=centering,labelsep=newline}
\caption{$Q$ value in the reaction 
$\, \Lambda_Q\Lambda_{\cal Q^\prime} 
\to \Xi_{\cal QQ^\prime}\, N$, $\,{\cal Q,Q^\prime}=s,c,b$.
\label{tab:Q}}
\begin{center}
\begin{tabular}{|c | r | r | r |r|} \hline \hline
  Observable (MeV)  & ${\cal Q,Q^\prime} = s$ 
   & ${\cal Q,Q^\prime}=c$\kern2em\strut 
 & ${\cal Q,Q^\prime}=b$  
\kern1em\strut & ${\cal Q}=b,{\cal Q^\prime}=c$
\\ \hline
$M(\Lambda_{\cal Q})$  & 1115.7     & 2286.5 \kern2.7em\strut
                                            & 5619.6\kern1.5em
& 5619.6, 2286.5
\\ 
\hline
\vrule width0pt height2.2ex
$M(\Xi_{\cal QQ^\prime})$   & 1314.9$^a$\kern-0.4em 
                                   & $\,\,\,3621.4\pm0.78$ 
& $10162\pm12^{\,b}$\kern-0.4em & $6917\pm13^{\,c}$\\
\hline\hline
$Q$-value             & ${-}23.1$  & ${+}12.0\pm0.78$   & ${+}138\pm12$  
& ${+}50\pm13$\kern0.4em\\ 
\hline\hline
\end{tabular}
\\
{\footnotesize
\vrule width 0pt height 3.0ex
$^a$To optimize the $Q$-value we take here $\Xi^0(ssu)$, 
$N{=}n$, because $M[\Xi^-(ssd)]$ is 7 MeV larger.
\\
$^b\,\Xi_{bb}$ mass prediction from Ref.~\cite{Karliner:2014gca}.
\kern 24.6em \strut
\\
\vskip-0.1cm
$^c$Average of the two values in Table XI of Ref.~\cite{Karliner:2014gca}.
\kern 18.0em \strut
}
\\
\end{center}
\end{table}
Clearly, the exothermic reaction \eqref{eq:LcLc} is the quark-level
analogue of the well-known exothermic nuclear fusion reactions
involving the lightest nuclei with two or three nucleons
\cite{nfusion},
\beq
\begin{array}{cccc}
\label{eq:DT}
D\,T&\to& {}^4{\rm He}\,n & \strut\kern8em\hbox{$Q=17.59$ MeV,}
\\
D\,D&\to&{}^3{\rm He}\,n & \strut\kern8em\hbox{$Q=\,\,3.27$ MeV,}
\\
D\,D&\to&T\,p & \strut\kern8em\hbox{$Q=\,\,4.04$ MeV,}
\\
T\,T&\to& {}^4{\rm He}\,2n & \strut\kern8em\hbox{$Q=11.33$ MeV,}
\\
D{}^3{\rm He}&\to&{}^4{\rm He}\,p & \strut\kern8em\hbox{$Q=18.35$ MeV,}
\\
{}^3{\rm He} {}^3{\rm He}&\to& {}^4{\rm He}\,2p & 
\strut\kern8em\hbox{$Q=12.86$ MeV.}
\end{array}
\eeq
It is interesting that the reaction \eqref{eq:LcLc} involves two {\em hadrons}
with three quarks each, rather than two nuclei with two or three nucleons
each, and that its $Q$-value is of similar order of magnitude
to the reactions~\eqref{eq:DT}.

Table~\ref{tab:Q} lists the $Q$-values of the four analogous reactions
$\,\Lambda_{\cal Q}\Lambda_{\cal Q^\prime} \to \Xi_{\cal QQ^\prime}\, N$, 
$\,{\cal Q,Q^\prime}=s,c,b$. 
The trend  is clear: 
the $Q$-values increase monotonically with increasing quark mass.
The reaction
\beq
\label{eq:LL}
\Lambda \Lambda \to \Xi N
\eeq
is {\em endothermic} with $Q={-}23$ MeV.
Reaction \eqref{eq:LcLc} is exothermic with $Q={+}12$ MeV, while the reaction
\beq
\label{eq:LbLb}
\Lambda_b \Lambda_b \to \Xi_{bb} N 
\eeq
is expected to be strongly exothermic with $Q={+}138\pm12$ MeV. 
Finally, the reaction
\beq
\label{eq:LbLc}
\Lambda_b \Lambda_c \to \Xi_{bc} N 
\eeq
is expected to have $Q=+50\pm13$ MeV, intermediate between $cc$ and $bb$.
The two latter estimates rely on the predictions of the $\Xi_{bb}$ 
     and $\Xi_{bc}$ masses in Ref.~\cite{Karliner:2014gca}.

\begin{figure}[t]
\begin{center}
\includegraphics[width=0.95\textwidth]{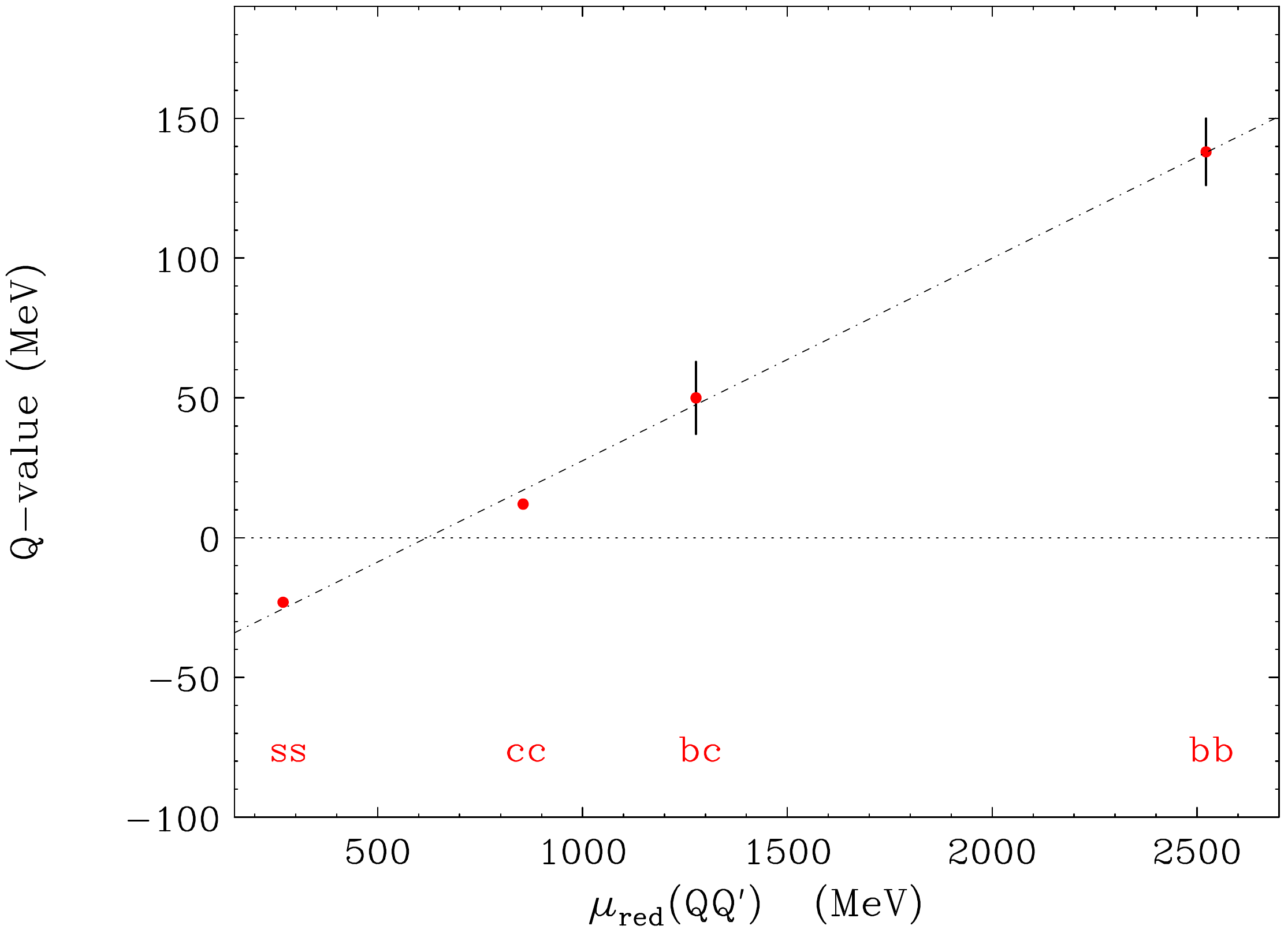}
\end{center}
\caption{\label{fig:QvsMU}
The $Q$-value in the quark-level fusion reactions
$\, \Lambda_Q\Lambda_{\cal Q^\prime} 
\to \Xi_{\cal QQ^\prime}\, N$, $\,{\cal Q,Q^\prime}=s,c,b$,
plotted against the reduced masses of the doubly-heavy diquarks 
$\mu_{red}({\cal Q Q^\prime})$.
The dot-dashed line denotes a linear fit $Q= -44.95 + 0.0726\,\mu_{red}$.}
\end{figure}

As already mentioned, the dominant effect determining the $Q$-value is
the binding between two heavy quarks. Since these quarks interact
through an effective two-body potential their binding is
determined by their reduced mass,
$\mu_{red}=m_{\cal Q} m_{\cal Q^\prime}/(m_{\cal Q} + m_{\cal Q^\prime})$.
In Figure~\ref{fig:QvsMU} we plot the $Q$-value 
vs.~$\mu_{red}({\cal Q Q^\prime})$.
The effective quark masses are taken as in Ref.~\cite{Karliner:2014gca}:
$m_s =538$ MeV, $m_c =1710.5$ MeV, $m_b =5043.5$ MeV.  
The straight line fit $Q= -44.95 + 0.0726\,\mu_{red}$ denoted by dot-dashed line
describes the data rather well, showing that to a good approximation
the $Q$-value indeed depends linearly on the reduced mass.

In addition to the reactions 
\eqref{eq:LcLc}, \eqref{eq:LbLb}, and \eqref{eq:LbLc} which 
involve fusion of two heavy {\em baryons} to a doubly-heavy baryon and a
nucleon, reactions involving fusion of two heavy {\em mesons} 
into a stable \cite{Karliner:2017qjm} doubly-heavy tetraquark 
$T(bb \bar u \bar d)$ are also possible, 
\beq
\label{eq:BBstar}
B^- \bar B^{0*}  \to T(bb \bar u \bar d), \qquad\qquad Q = 215\hbox{\ MeV},
\eeq
and
\beq
\label{eq:BB}
B^- \bar B^0  \to T(bb \bar u \bar d)\,\gamma, 
\qquad\qquad Q = 170\hbox{\ MeV}.
\eeq
Reaction \eqref{eq:BBstar} is analogous to fusion of proton and neutron 
into deuteron, with $Q=2.2$ MeV. Reaction \eqref{eq:BB} has a lower
$Q$-value than \eqref{eq:BBstar} and it requires EM interaction 
on top of QCD, in order to conserve angular momentum and parity, 
since $T(bb \bar u \bar d)$ has $J^P=1^+$ \cite{Karliner:2017qjm}. 

It is worth noting that Table~\ref{tab:Q} anticipates a 
strong violation of the would be heavy-quark analogue of the 
Gell-Mann--Okubo mass formula \cite{GellMann:1961ky,Okubo:1961jc}
\beq
\label{eq:GMO}
\frac{N+\Xi}2 = \frac{3\Lambda + \Sigma}4
\eeq
where l.h.s.$\,=1128.6$ MeV and r.h.s.$\,=1135.1$ MeV, i.e.,
Eq.~\eqref{eq:GMO} is accurate to $\sim0.6\%$.
Indeed, for charm l.h.s.$\,=2280$ MeV and r.h.s.$\,=2328$ MeV, while
for bottom l.h.s.\,=$5551\pm6$ MeV and r.h.s.$\,=5668$ MeV.

One might think that this is not surprising, given that the
Gell-Mann--Okubo mass formula was derived assuming a small breaking of flavor
SU(3) and there is no corresponding flavor symmetry when the $s$-quark is
replaced by $c$ or $b$. But there is more to it than this. 

In the modern view Eq.~\eqref{eq:GMO} results from equal numbers of light
and $s$ quarks on both sides of the equation, together with the
corresponding spin-dependent color-hyperfine interaction terms.
Had this been the whole story, the mass formula \eqref{eq:GMO} should have 
approximately worked also for $c$ and $b$. The reason it fails is that
the large binding energy between the two heavy quarks is not included in the 
derivation.

It should be stressed that from the point of view of the strong interactions,
all the $\Lambda_{\cal Q}$ and $\Xi_{\cal QQ^\prime}$ baryons
are {\em stable particles}. They eventually do decay, via weak
interactions, but their lifetimes are at least ten orders of magnitude
longer than the typical timescale $10^{-23}$~s of the fusion reactions 
\eqref{eq:LcLc}, \eqref{eq:LbLb}, and \eqref{eq:LbLc}
which proceed purely through strong interactions.
An analogous observation applies to reaction \eqref{eq:BBstar}.

\medskip
\noindent
{\bf Analogy with doubly strange hypernuclei}
\nl
Implications of the strong binding of two heavy quarks in $\Xi_{\cal QQ}$ 
go beyond the fusion-like reactions
$\,\Lambda_{\cal Q}\Lambda_{\cal Q} \to \Xi_{\cal QQ}\, N,\,\,{\cal Q}=b,c$. 
There might also be interesting ramifications for $cc$ and $bb$ analogues 
of hypernuclei.

Ref.~\cite{Harada:2010vd}
examined theoretically the production of doubly strange hypernuclei, 
$_{\Xi^-}^{\,\,16}{\rm C}$ and $_{\Lambda\Lambda}^{\,\,16}{\rm C}$
in double-charge exchange $^{16}{\rm O}(K^-, K^+)$ reactions,
where $_{\Xi^-}^{\,\,16}{\rm C}$ denotes a $Z=6$ hypernucleus with 
$\Xi^-$ in place one of the original nucleons.

We conjecture that in principle an analogous reaction with $b$
instead of $s$-quarks might be possible, i.e.,
$^{16}{\rm O}(B^-, B^+)^{16}_{\Xi_{bb}^-}{\rm C}$,
or
\beq
\label{eq:BOOB}
B^-\, {}^{\,\,16}{\rm O} \to B^+\,{}^{\,\,16}_{\Xi_{bb}^-}{\rm C}\,.
\eeq
The difference between
the binding energy of $^{\,\,16}{\rm O}=7.98$ MeV and
of ordinary $^{\,\,16}{\rm C}=6.92$ MeV \cite{nuclides} is only about 1 MeV.
On the other hand, the binding energy of $bb$ in $\Xi_{bb}$ is $\sim 280$
MeV, so the reaction \eqref{eq:BOOB} is expected to have a rather large
$Q$-value.

Experimentally this reaction is extremely challenging, because the $B^-$
lifetime is only $1.6\times 10^{-12}$~s, four orders of magnitude shorter
than $\tau(K^-) = 1.2\times 10^{-8}$ s.  To put it in perspective, the
distance $d_B$ covered by $B^-$ is $d_B = \gamma(B) \cdot \tau(B^-)\cdot c
= \gamma(B)\cdot 480\,\mu$, where $\gamma(B^-)$ is the Lorentz factor. So
a 10 GeV $B^-$ will travel $\sim 1$ mm before decaying.

An analogous reaction with a $c$ instead of a $b$-quark might perhaps also 
be possible, i.e.,
\beq
\label{eq:DAAD}
D^+\, {}^{N}\kern-0.2em{\rm A} \to D^-\,{}_{\Xi_{cc}^{++}}^{\kern0.8em N}
\kern-0.1em{\rm A}\kern-0.1em{}^\prime\,. 
\eeq
Such a reaction could take place in a collision in which at least one of the
two projectiles is a heavy ion.  The initial $D^+$ would be produced as part
of an open-charm process, and would then undergo double-charm-exchange in the
target nucleus.

\section*{ACKNOWLEDGMENTS}
The work of
J.L.R. was supported in part by the U.S. Department of Energy, Division of
High Energy Physics, Grant No.\ DE-FG02-13ER41958.

\end{document}